\documentclass[twocolumn,showpacs,preprintnumbers,floats,aps,pra,nofootinbib,10pt]{revtex4-1}
\usepackage{graphicx}
\usepackage{dcolumn}
\usepackage{amsmath}
\usepackage{amsfonts}
\usepackage{amssymb}
\usepackage{dcolumn}% Align table columns on decimal point
\usepackage{color}
\usepackage{float}
\usepackage{subfig}
\usepackage{tabularx}
\usepackage{soul}
\usepackage{ulem}

\newcolumntype{L}[1]{>{\raggedright\arraybackslash}p{#1}}
\newcolumntype{C}[1]{>{\centering\arraybackslash}p{#1}}
\newcolumntype{R}[1]{>{\raggedleft\arraybackslash}p{#1}}

\definecolor{v}{rgb}{0.6, 0.2, 0.8} %comentarios VM
\providecommand{\U}[1]{\protect\rule{.1in}{.1in}}

\begin{document}
\title{Stability analysis and constraints on interacting viscous cosmology}

\author{A. Hern\'andez-Almada$^1$}
\email{ahalmada@uaq.mx}

\author{Miguel A. Garc\'ia-Aspeitia$^{2,3}$}
%\email{aspeitia@fisica.uaz.edu.mx}

\author{Juan Maga\~na$^{4,5}$}
%\email{juan.magana@uv.cl}

\author{V. Motta$^6$}
%\email{veronica.motta@uv.cl}

\affiliation{$^1$Facultad de Ingenier\'ia, Universidad Aut\'onoma de Quer\'etaro, Centro Universitario Cerro de las Campanas, 76010, Santiago de Quer\'etaro, M\'exico}

\affiliation{$^2$Unidad Acad\'emica de F\'isica, Universidad Aut\'onoma de Zacatecas, Calzada Solidaridad esquina con Paseo a la Bufa S/N C.P. 98060, Zacatecas, M\'exico.}
\affiliation{$^3$Consejo Nacional de Ciencia y Tecnolog\'ia, \\ Av. Insurgentes Sur 1582. Colonia Cr\'edito Constructor, Del. Benito Ju\'arez C.P. 03940, Ciudad de M\'exico, M\'exico.}

\affiliation{$^4$Instituto de Astrof\'isica, Pontificia Universidad Cat\'olica de Chile, Av. Vicu\~na Mackenna, 4860, Santiago, Chile.}
\affiliation{$^5$Centro de Astro-Ingenier\'ia, Pontificia Universidad Cat\'olica de Chile, Av. Vicu\~na Mackenna, 4860, Santiago, Chile.}
\affiliation{$^6$Instituto de F\'isica y Astronom\'ia, Facultad de Ciencias, Universidad de Valpara\'iso, Avda. Gran Breta\~na 1111, Valpara\'iso, Chile.}

\begin{abstract}
In this work we study the evolution of a spatially flat Universe by considering a viscous dark matter and perfect fluids for dark energy and radiation, including an interaction term between dark matter and dark energy. In the first part, we analyse the general properties of the Universe by performing a stability analysis and then we constrain the free parameters of the model using the latest and cosmological-independent measurements of the Hubble parameter. We find consistency between the  viscosity coefficient and the condition imposed by the second law of the Thermodynamics. The second part is dedicated to constrain the free parameter of the interacting viscous model (IVM) for three particular cases: the viscous model (VM), interacting model (IM), and the perfect fluid case (the concordance model). 
We report the deceleration parameter to be 
$q_0 = -0.54^{+0.06}_{-0.05}$, 
$-0.58^{+0.05}_{-0.04}$, 
$-0.58^{+0.05}_{-0.05}$, 
$-0.63^{+0.02}_{-0.02}$, together with the jerk parameter as
$j_0 = 0.87^{+0.06}_{-0.09}$, 
$0.94^{+0.04}_{-0.06}$, 
$0.91^{+0.06}_{-0.10}$, 
$1.0$ for the IVM, VM, IM, and LCDM respectively, where the uncertainties correspond at 68\% CL. Worth mentioning that all the particular cases are in good agreement with LCDM, in some cases producing even better fits, with the advantage of eliminating some problems that afflicts the standard cosmological model. 
\end{abstract}

%\draft
\date{\today}
\maketitle

%%%%%%%%%%%%%%%%%%%
\section{Introduction}
%%%%%%%%%%%%%%%%%%%

Dark energy (DE) and dark matter (DM) are the corner stones of the modern cosmology, being so far one of the most intriguing mysteries for the understanding of our universe. In this vein, many attempts to comprehend the composition of this dark entities have been developed in recent years. The most important theories for DM are supersymmetry models \citep{Martin:1997}, scalar fields \cite{Magana:2012ph,Matos:1999et,Hernandez-Almada:2017mtm}, interacting dark energy \cite{PhysRevD.99.043521,VONMARTTENS:2019, VONMARTTENS:2020}, charged particles coming from unbroken $U(1)$ gauge symmetry featuring dissipative interactions \cite{Vagnozzi:2015}, among others, meanwhile for DE the most interesting candidates can be summarized as the cosmological constant (CC), phantom energy, quintessence, Chaplygin gas, brane-worlds, f(R), unimodular gravity etc. (see Refs. \citep{Copeland:2006, Bamba:2012rev} for some reviews of DE models, also see \cite{Garcia-Aspeitia:2016kak,Garcia-Aspeitia:2018fvw,Hernandez-Almada:2018osh,Garcia-Aspeitia:2019yni,Astorga-Moreno:2019uin,Garcia-Aspeitia:2019yod}). Despite the efforts of the community, the supersymetric DM and CC as dark energy are still the best candidates to understand the cosmological observations. However, laboratory experiments show no evidence of supersymmetric particles and the CC is afflicted with several theoretical problems \citep{Weinberg,Zeldovich} when its origin is considered as quantum vacuum fluctuations. A radical new form to address these conflicts is to consider an interaction between the dark components through the continuity equation \cite{Chimento:2000, Copeland:2006, Bolotin:2015}.

On the other hand, cosmology with viscous dark fluids is an interesting alternative to understand the accelerated expansion of the Universe \cite{Fabris:2006}. The viscous fluid models could resolve the tension across different probes, for instance, the value of the Hubble constant ($H_0$) obtained from SNIa \cite{Anand:2017, Riess:2016} is more than $3\sigma$ the one estimated by CMB Planck data \cite{Planck:2016rXLVI}, and the value of matter fluctuation amplitude ($\sigma_8$) measured from the large scale structure (LSS) observations differs from those determined from the CMB Planck data under the LCDM cosmology \cite{DES:2018v26, DES:2018v28}. Authors in \cite{Kremer:2012} study a dissipative Universe with interacting fluids which the non-equilibrium pressure is proportional to $H_0$, they find that the decelerated - accelerated transition occurs earlier than the one of a non-viscous model (for other interesting models, see \cite{Avelino:2013, Atreya:2018, Valentino2019}).

Although there are two types of viscosity coefficient known as bulk and shear, the bulk viscosity is the one that plays an important role in the Universe's dynamics at the background level because it satisfies the cosmological principle. In contrast, one of the main characteristics of the shear viscosity is that it could produce vortices or any other chaotic phenomena at early epochs of the Universe evolution. 
Based on bulk viscosity, the viscous models have been addressed using two approaches: the Eckart \cite{Eckart:1940} and Israel-Stewart (IS) \cite{Israel1979} theories. For an extensive review on viscous cosmology, see \cite{Brevik:2017rev}. The main difference between the theories is that the IS approach explored by \cite{Zimdahl:1996, Mak:1998, Paul:1998} solves {\it the problem of the causality}, i.e., the propagation of the perturbations on the viscous fluids is superluminal. Although this formalism avoids this problem is more complex than the Eckart theory, and only some analytical solutions \cite{MCruz:2017, NCruz:2018, NCruz:2018arx, NCruz:2019} are known for the bulk viscosity of the form $\xi \sim \xi_0 \rho^s$, with $s=1/2$ and $\rho$ is the energy density of the viscous fluid in an Universe filled by only one fluid \cite{Burd:1988}.  In contrast, the Eckart's scenario was the first proposal to study the relativistic dissipative processes as a first order deviations around the equilibrium and, despite the causality problem, it is widely used due to its simplicity. For instance, some works related to the Eckart's theory, have been investigated the dynamics of the Universe at late times by considering a bulk viscous coefficient with a constant \cite{Murphy:1973, Padmanabhan:1987, Brevik2005, Normann:2017}, polynomial \cite{Xin-He:2009, Avelino:2010, Almada:2019}, and hyperbolic \cite{FOLOMEEV200875, Almada:2019} forms as functions of the redshift or in terms of the energy density. Additionally, authors in \cite{Normann:2016,Normann:2017} have been studied the Universe with several fluids, being a more realistic description of the Universe.
In both theories, the procedure to include the bulk viscous effects into the Einstein field equations is trough as an effective pressure, written in the form $\tilde{p} = p + \Pi$, where $p$ refers to the sum of the traditional components such as the dust-matter (baryons, DM), the DE, and the relativistic species (photons, neutrinos), being $\Pi$ an account to the bulk viscosity term. As a consequence, the equation of state (EoS) generally turns into an inhomogeneous one when the $\Pi$ term is a variable function. 
Furthermore, it is worth noticing that letting $\Pi$, or any other dynamical variable, vary with time is an interesting way to explain the recent results given by \cite{Zhao:2017}, which conclude a  preference of the DE component for a dynamical EoS over a constant one.
Regarding the physical mechanism to generate such viscous effects, some proposals point towards the decaying of DM particles \cite{Wilson:2007, Mathews:2008} or any other microscopic property as the self-interaction \cite{Atreya:2018} of DM particles.

Recently, the Experiment to Detect the Global EoR Signature (EDGES) \cite{Bowman:2018} found that the amplitude of the absorption signal of 21 cm temperature at the cosmic dawn epoch ($z\approx 17$) is larger than expected. In this vein, the EDGES observations indicate that the baryons must be cooler or the photons hotter than the predicted by the standard cosmology, thus, this phenomenon offers another incentive to study the viscosity effects of the fluids and their interactions \cite{Barkana:2018, Bhatt:2019}. Considering the first and second law of Thermodynamics and assuming the Universe filled by a non perfect DM fluid with $\xi \sim \rho^s$, the authors in \cite{Bhatt:2019} find that the temperature of the DM fluid increase throughout the cosmic evolution due the bulk viscosity $\xi_0>0$, thus, allow to describe the EDGES observations.

Therefore, in this work we study a model that consists of a flat Friedmann-Lemaıtre-Robertson-Walker (FLRW) Universe including three components: a non-perfect and interacting fluid, composed by DM where baryons are included, which we will call it as dust matter (dm) \footnote{Other models in literature separates the baryons from dark matter.}, the DE fluid that will interact with dm  in the Eckart's approach and radiation with its standard well known behavior. We start analyzing the general dynamics of these components through a stability analysis of the critical points. After that, we perform a Monte Carlo Chain Markov (MCMC) procedure using the latest observational Hubble parameter data (OHD) to constrain the free parameters of the model\footnote{ For instance, see \cite{Biswas:2005wy} as another alternative to perform the dynamical system analysis in combination with Bayesian MCMC analysis.}. Finally, we study particular cases of the model such as a solely viscous model (without the interaction term), an interacting model (without the viscosity term), and the perfect fluid case that correspond to the LCDM model.

The paper is organized as follow: In Sec. \ref{sec:CDF} presents the background of the interacting non-perfect model and gives the formulation of the dynamical system. In Sec. \ref{sec:SA} we discuss the stability of the system around the critical points and give bounds to the free model parameters. Section \ref{sec:data} is devoted to constrain the free parameters of the model using the latest samples of OHD. In Sec. \ref{sec:Res} we discuss our results and finally, we present our remarks and conclusions in Sec.\ref{sec:Con}.

%%%%%%%%%%%%%%%%%%%%%%%%%%%%%%%%%%%%%%%%%
\section{Cosmology with Dark Fluids} \label{sec:CDF}
%%%%%%%%%%%%%%%%%%%%%%%%%%%%%%%%%%%%%%%%%

The cosmological model under study consists of a Universe in a flat FLRW space time which contains a non-perfect fluid as dm that interacts with a perfect fluid as the DE component, together with the radiation fluid. Then, the energy-momentum tensor can be expressed as
\begin{equation}
    T_{\mu\nu} = \rho u_{\mu}u_{\nu} + \tilde{p}(g_{\mu\nu} + u_{\mu}u_{\nu})
\end{equation}
where $g_{\mu\nu}$ corresponds to the FLRW metric, $\tilde{p}=p+\Pi$ is the sum of the total barotropic pressure of the fluids, $p$, and the bulk viscosity coefficient, $\Pi$, $\rho$ is the energy density of the fluid and $u_{\mu}$ is the associated cuadri-velocity. Inspired on the viscosity behavior in fluid mechanics, being proportional to the speed, we have assumed $\Pi=-3 \zeta H$. Additionally, the model supposes an energy exchange term $Q$ between dm and DE, and a viscosity effect encoded in the terms that contain the bulk viscosity coefficient $\zeta$.
In this approach, the Friedmann, continuity and acceleration equations are
\begin{eqnarray}
    &&H^2=\frac{\kappa^2}{3}\left( \rho_r + \rho_{dm} + \rho_{de} \right), \label{eq:VFEa} \\
    &&\dot{\rho}_{r} + 4 H \rho_r = 0 \,,\label{eq:VFEb} \\
    &&\dot{\rho}_{dm} + 3 H \rho_{dm} = 9 H^2\zeta + Q \,, \label{eq:VFEc}\\
    &&\dot{\rho}_{de} + 3 \gamma_{de} H \rho_{de} = - Q \,, \label{eq:VFEd}\\
    &&2\dot{H} - 3 \kappa^2 H \zeta = -\kappa^2 \left( \rho_{dm} + \frac{4}{3}\rho_{r} + \gamma_{de}\rho_{de} \right) \,, \label{eq:VFEe}
\end{eqnarray}
where $H=\dot{a}/a$, $\kappa^2 = 8\pi G$, $G$ is the Newton gravitational constant, $\rho_r$, $\rho_{dm}$, and $\rho_{de}$ correspond to the relativistic species, dust matter and dark energy  densities respectively. The equation of state (EoS) for each species are $p_{r}=\rho_r/3$, $p_{dm}=0$, and $p_{de} = (\gamma_{de}-1)\rho_{de}$, being $\gamma_{de}$ a constant that it is related with the EoS as $\omega_{de}=\gamma_{de}-1$. Notice that the DE component behaves as CC when $\gamma_{de}=0$. 
 
In particular, in this work we consider the typical ansatz for the viscosity coefficient
\begin{equation}
    \zeta = \frac{\xi}{\kappa^2} \left( \frac{\rho_{dm}}{\rho_{dm0}} \right)^{1/2} \,,
\end{equation}
where $\rho_{dm0}$ is the dm density at present epoch and $\xi$ is a free parameter with units of $[\xi]=$[eV].

To study the cosmological model presented in Eqs. (\ref{eq:VFEa})-(\ref{eq:VFEe}), we define the dimensionless dynamical variables as 
\begin{equation}\label{eq:vars}
    x= \frac{\kappa^2\rho_{de}}{3H^2}\,,\, y= \frac{\kappa^2\rho_{dm}}{3H^2}\,,\, \Omega_{r} = \frac{\kappa^2\rho_{r}}{3H^2}\,,\, z=\frac{\kappa^2 Q}{3H^3}\,.
\end{equation}
From Eq. (\ref{eq:VFEa}), it is straightforward to see that $\Omega_r= 1 - x - y$. Then, the dynamical system can be written as \cite{Leyva:2017}
\begin{eqnarray}
    x' &=& 3(x-1)x\gamma_{de} - 3\xi_0 x y^{1/2} - x( 4x + y -4 ) \nonumber \\
       & & - z(x,y)\,, \label{eq:sd1}\\
    y' &=& 3\gamma_{de}xy - y(4x+y-1) - 3 \xi_0 (y-1)y^{1/2} \nonumber \\ 
       & & + z(x,y)  \,, \label{eq:sd2}
\end{eqnarray}
where $'= d/dN$, $N = \log(a)$ and
\begin{equation}
    \xi_0 = \frac{\xi}{H_0 y_0^{1/2}}\,.
\end{equation}
In the latter equation notice that $y_0$  and $H_0$ are the fraction of dm and Hubble parameter at $z=0$ respectively. Additionally, to convert the dynamical system in an autonomous one we have defined the variable $z$ related to the interaction term. In particular, we will explore the form of $Q$ as
\begin{equation}
    Q = \beta H \frac{\rho_{de}\rho_{dm}}{\rho_{de}+\rho_{dm}}\,,
\end{equation}
or in terms of the dimensionless variables \cite{Zimdahl:2003}
\begin{equation} \label{eq:z}
    z(x,y) = \beta \frac{xy}{x+y}\,,
\end{equation}
where $\beta$ is a dimensionless free parameter. It is evident that, for $\beta=0$, the system described above corresponds to an Universe with viscosity. 
For alternative forms of $z(x,y)$ that satisfy such conditions, see for instance \cite{Leyva:2017}. 
In addition, we express the deceleration parameter, effective EoS and jerk parameter as \cite{Leyva:2017}
\begin{eqnarray}
    q(N)       &=& 1 -\left( 2 - \frac{3}{2}\gamma_{de} \right) x - \frac{1}{2}y - \frac{3}{2} \xi_0 y^{1/2} \,, \\
    w_{eff}(N) &=& \frac{1}{3}\left[ 1 - (4-3\gamma_{de})x -y - 3\xi_0 y^{1/2} \right] \,, \\
    j(N)       &=& q (2q+1) - q^{\prime} \,,
\end{eqnarray}
where previous equations are written in terms of the dimensionless variables.

%%%%%%%%%%%%%%%%%%%%%%%%%%%%%%%%%%%%%%%%%%%%%
\section{Stability analysis} \label{sec:SA}
%%%%%%%%%%%%%%%%%%%%%%%%%%%%%%%%%%%%%%%%%%%%%

We start our stability study of the dynamical variables defined in the Eqs. (\ref{eq:vars})-(\ref{eq:sd2}) by finding the critical points and the Jacobian matrix, which are respectively
\begin{eqnarray}
P_1 = (0\,,\, 0)\,,\;\; P_2 = (0\,,\, 1)\,,\;\; P_3 = (1\,,\, 0)\,, \label{eq:EqP} 
\end{eqnarray}
and 
\begin{equation}
J =
\begin{pmatrix}
J_{xx} &  J_{xy} \\
J_{yx} &  J_{yy}  
\end{pmatrix},
\end{equation}
where
\begin{eqnarray}
    J_{xx} &=&  4 -8x - y -3 \xi_0 y^{1/2} - \beta \frac{y}{x + y} \nonumber \\
           & &  + \beta \frac{xy}{(x + y)^2}, \\
    J_{xy} &=& - x -\frac{3}{2} \xi_0 xy^{-1/2} - \beta \frac{x}{x + y} + \beta \frac{xy}{(x + y)^2},   \\
    J_{yx} &=& - 4y + \beta \frac{y}{x + y}  -\beta \frac{xy}{(x + y)^2},  \\
    J_{yy} &=& 1 - 4x - 2y -\frac{9}{2} \xi_0 y^{1/2} + \frac{3}{2} \xi_0 y^{-1/2}  \nonumber \\
           & & + \beta \frac{x}{x + y} - \beta \frac{xy}{(x + y)^2} \,.
\end{eqnarray}

\begin{table*}
\caption{Critical points and stability conditions for the IVM.}
\centering
\begin{tabular}{| C{3cm}C{3cm}C{3cm}C{5cm}  |}
\hline
Critical point   & ( $x\;,\;y$ )  & Eigenvalues                     &  Stability condition ($\Re(\lambda)<0$)               \\
\hline
$P_1$            & ( $0\;,\;0$ )     & $4-\beta\,,\; \infty  $          &   Saddle if $\beta>4$ \\ [0.7ex]
$P_2$            & ( $0\;,\;1$ )     & $3-3\xi_0 -\beta\,,\; -1-3\xi_0$ &   $\beta>3(1-\xi_0)$ and $\xi_0> -\frac{1}{3}$\\ [0.7ex]
$P_3$            & ( $1\;,\;0$ )     & $-4\,,\; \infty$                 &   Saddle \\ [0.7ex]
\hline
\end{tabular}
\label{tab:CP}
\end{table*}

The stability analysis of non linear systems consists in studying the behavior of the perturbations around the critical points using the matrix $J$ and decide if they are stable or not. 
Notice that for a vector $\vec{x}=(x,y,\Omega_r,z)$ that contains all the dynamical variables described in Eq. (8), we considered a small perturbation $\vec{x}\to \vec{s}+\delta \vec{x}$ around the critical (or equilibrium) point $s_i$, thus, an associated system is obtained in the form $\delta\vec{x}^{\prime}=J_{s_i}\delta\vec{x}$, where $J$ is the previously mentioned Jacobian matrix at the point $s_i$. Hence, the Hartman-Grobman theorem guarantees that, for a critical point, there exists a neighborhood for which the flow of the system of dynamical equations is topologically equivalent to the linearized one \cite{Grobman,Hartman} (see also \cite{Coley:2003,Leon:2014} for the dynamical system analysis in Cosmology). Hence, Table \ref{tab:CP} summarizes the stability condition for each critical point. The first point, $P_1 = (0,0)$, represents the radiation dominant epoch with $q=1$ and $w_{eff} = 1/3$. Notice that this point is a saddle for $\beta>4$ and unstable for $\beta<4$. The latter condition guarantees the evolution of the Universe to another critical point that is expected to be $P_2$.

\begin{figure*}
  \centering
  \includegraphics[width=.32\linewidth]{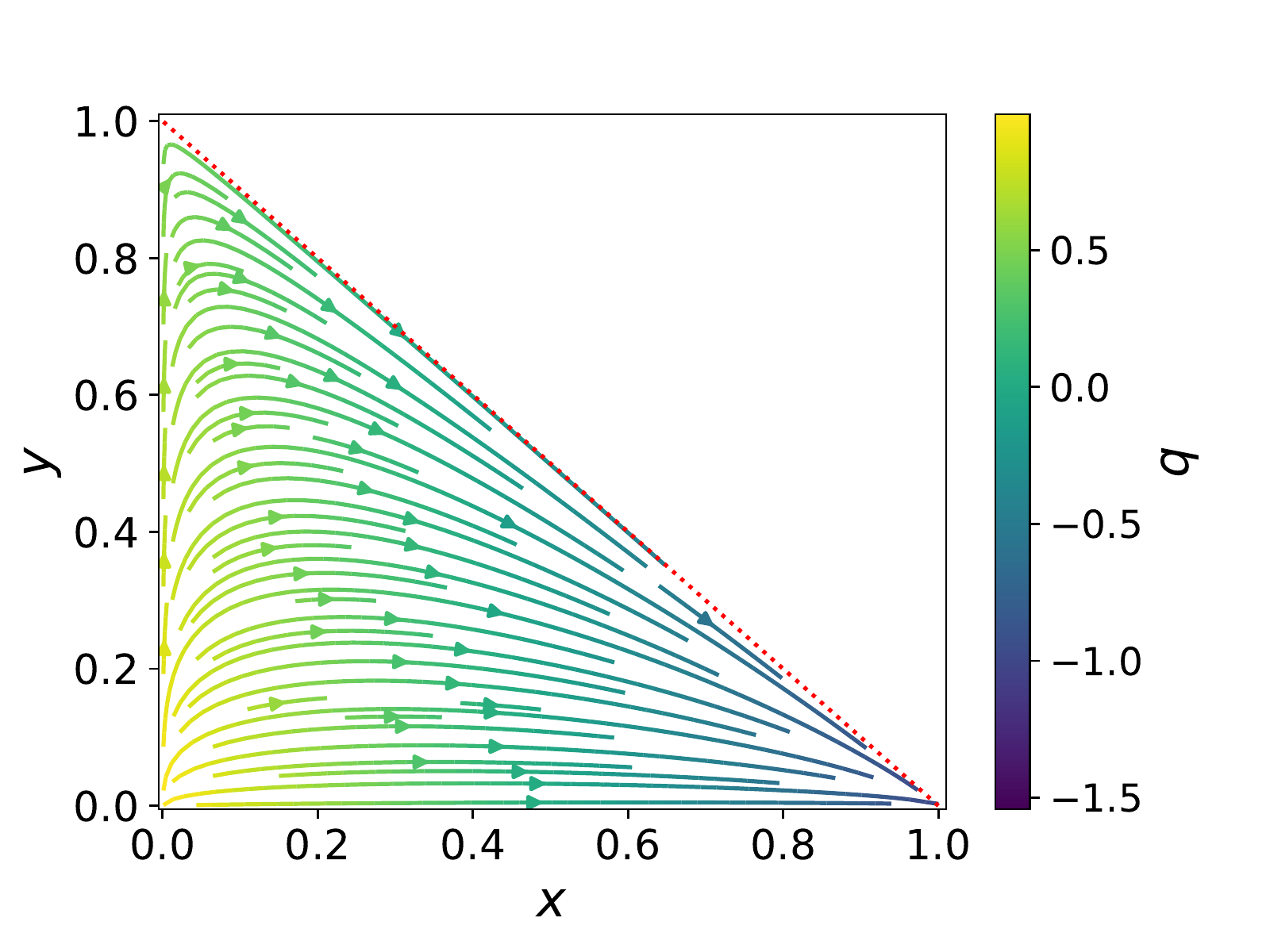} 
  \includegraphics[width=.32\linewidth]{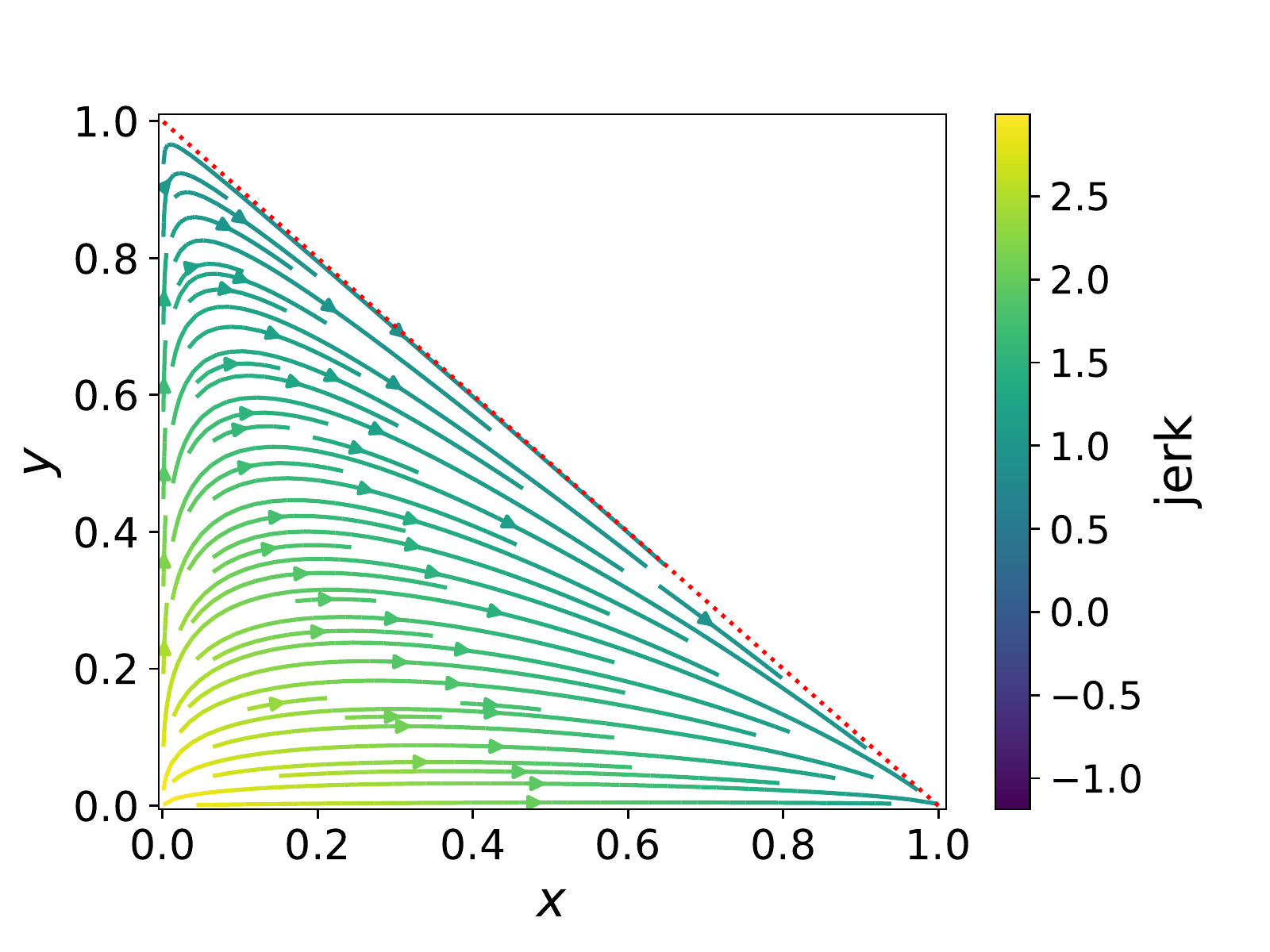} 
  \includegraphics[width=.32\linewidth]{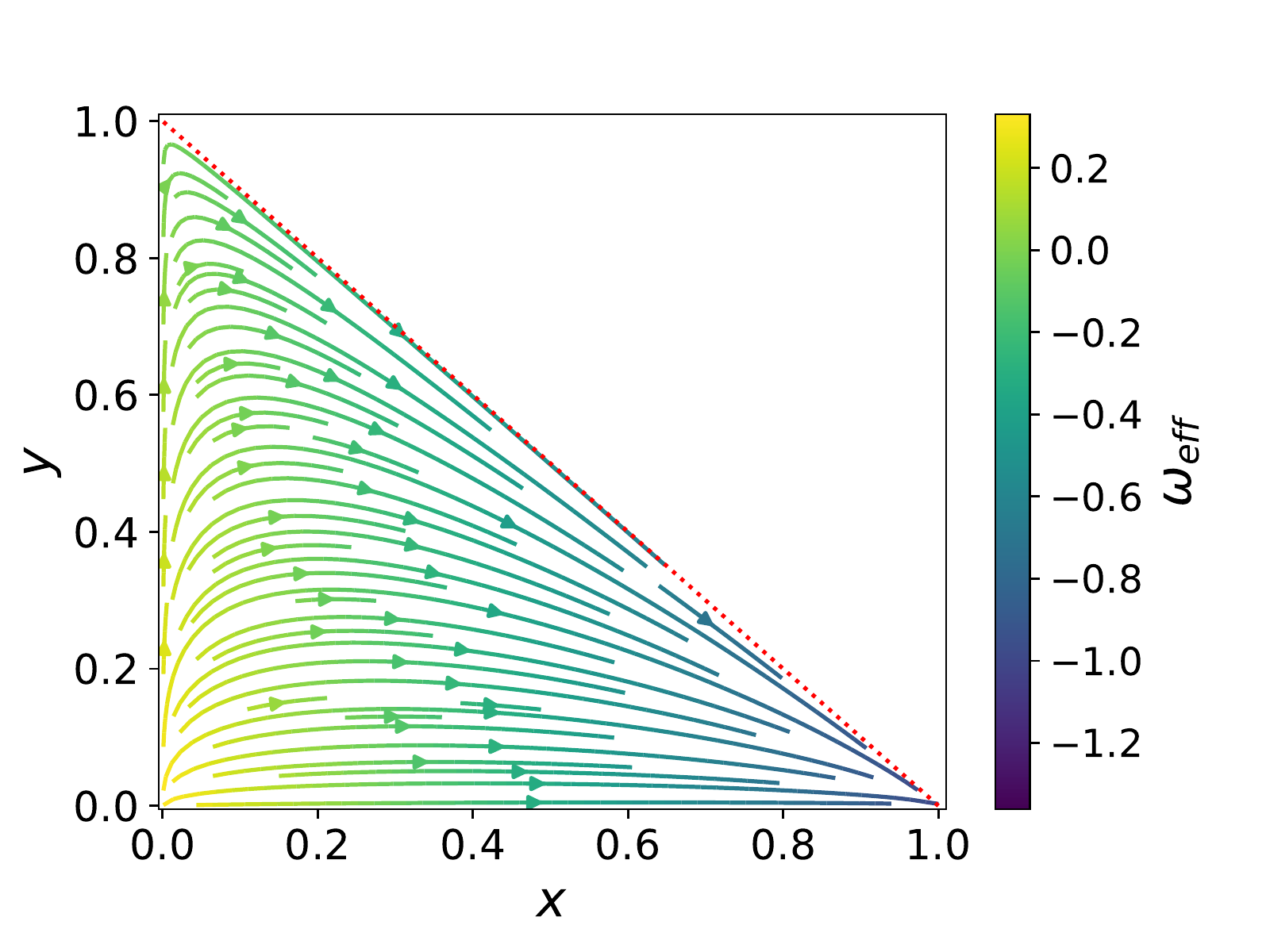} 
\caption{\{x,y\}-phase space using $h=0.701$, $\Omega_{de}=0.682$, $\gamma_{de} = 0$, $\beta = 0.200$ and $\xi_0 = 0.028$ (see Sec. \ref{sec:data} for details). On the left (middle, right) panel, the bar color represents the value of the deceleration (jerk, effective EoS) parameters. The diagonal red dotted line is the curve $x + y = 1$ for the case $\Omega_{r}=0$.} 
\label{fig:phasespace}
\end{figure*}

The $P_2$ point corresponds to the DM dominant epoch and it is stable in the region $\beta>3(1-\xi_0)$ and $\xi_0> -1/3$. On the other hand, $P_3$ is a saddle point if  $\beta<3(1-\xi_0)$ and  $\xi_0> -1/3$ and an unstable point if $\beta<3(1-\xi_0)$ and $\xi_0< -1/3$. Notice that the latter condition does not satisfy the second law of the thermodynamics that imposes $\xi_0>0$ \cite{Zimdahl:2000, Maartens:1996}, and the saddle point gives a weaker condition for $\xi_0$ than the thermodynamic one. Moreover, structure formation is explained by the existence of $P_2$ in our dynamical system where the interaction term has no contribution. Furthermore, Figure \ref{fig:phasespace} shows the \{x,y\}-phase space representing in gradient color the intensity of the deceleration (left panel) and jerk (middle panel) parameters and the effective EoS (right panel). In this phase-space, the evolution of the Universe starts around the point $P_1$ with $q\approx 1$, $j\approx 3$, and an effective EoS $w_{eff}\approx 1/3$. Then, depending on the initial conditions of the Universe, it could change to a state close to $P_2$ with cosmographic parameters $q\approx 1/2$, $j\approx 1$, and $w_{eff}\approx 0$. As mentioned before, this phase plays an important role in the structure formation, for that reason, an evolution with $y=0$ should not be allowed physically. The last stage of the Universe, where the accelerated expansion occurs, is when it moves towards the point $P_3$ with cosmographic parameters $q\approx -1/2$, $j\lesssim 1$, and $w_{eff}\approx -0.7$.

%%%%%%%%%%%%%%%%%%%%%%%%%%%%%%%%%%%%%%%%%
\section{Observational constraints} \label{sec:data}
%%%%%%%%%%%%%%%%%%%%%%%%%%%%%%%%%%%%%%%%%

The expansion rate of the Universe is measured directly by the OHD. Currently, the OHD sample is obtained from the differential age technique and Baryon Acoustic Oscillations (BAO) measurements. In this work we consider the sample compiled by \cite{Magana:2018}, that consists of  $51$ points in the redshift region $0.07<z<2.36$, to constrain the free model parameters. It is worth to note that this sample can yield biased constraints because the BAO points are estimated under a fiducial cosmology \cite{Magana:2018}. Thus, the figure-of-merit is given by
\begin{equation}\label{eq:chi2_ohd}
\chi^2_{OHD} = \sum_{i=1}^{51} \left( \frac{H_{th}(z_i, {\bf \Theta}) - H_{obs}(z_i)}{\sigma_{obs}^i} \right)^2 \,,
\end{equation}
where $H_{th}(z_i, {\bf \Theta})-H_{obs}(z_i)$ denotes the difference between the theoretical Hubble parameter with parameter space ${\bf \Theta}$ and the observational one at the redshift $z_i$, and $\sigma_{obs}^i$ is the uncertainty of $H_{obs}^i$.

The data will be used not only to constrain our interacting viscous model (IVM) with free parameter space ${\bf \Theta} = (h, \Omega_{de0}, \xi_0  ,\beta)$, but also the following particular models of the IVM: an only interacting model (IM) by setting $\xi_0=0$, an only viscous model (VM) with $\beta=0$, and the LCDM model that is recovered by requiring $\xi_0=0$ and $\beta=0$. In order to solve the equation system, Eqs. (\ref{eq:vars})-(\ref{eq:z}), we have used $\Omega_{de0}$ for the initial condition of $x$, and $y_0=1-\Omega_{de0}-\Omega_{r}$ for $y$, where $\Omega_r=2.469 \times 10^{-5}h^{-2} (1+0.2271N_{eff})$, with $N_{eff}=3.04$ as the number of relativistic species \cite{Komatsu:2011}, and $h$ as the Hubble dimensionless parameter. To minimize the $\chi^2$-function for each model, we perform a Bayesian MCMC analysis based on \textit{emcee} module \cite{Emcee:2013}. For each free model parameter, the n-burn phase is stopped following the Gelman-Rubin criteria \cite{Gelman:1992}, i.e. after achieving a value lower than 1.1.
We obtain $5000$ chains, each one with $500$ steps, to explore the confidence region taking into account a Gaussian prior on the Hubble constant $h$ and a flat prior for the rest of the parameters (see Table \ref{tab:priors}).

\begin{table}
\caption{Priors used in the MCMC analysis.}
\centering
\begin{tabular}{| C{3cm}C{4cm}  |}
\hline
Parameter       &  Prior                  \\
\hline
$h$             & Gauss$(0.7324,0.0174)$ \\ [0.7ex]
$\Omega_{de0}$  & Flat in $[0,1]$     \\ [0.7ex]
$\xi_0$         & Flat in $[0,1]$     \\ [0.7ex]
$\beta$         & Flat in $[0,3]$     \\ [0.7ex]
\hline
\end{tabular}
\label{tab:priors}
\end{table}

\begin{figure}
  \centering
  \includegraphics[width=\linewidth]{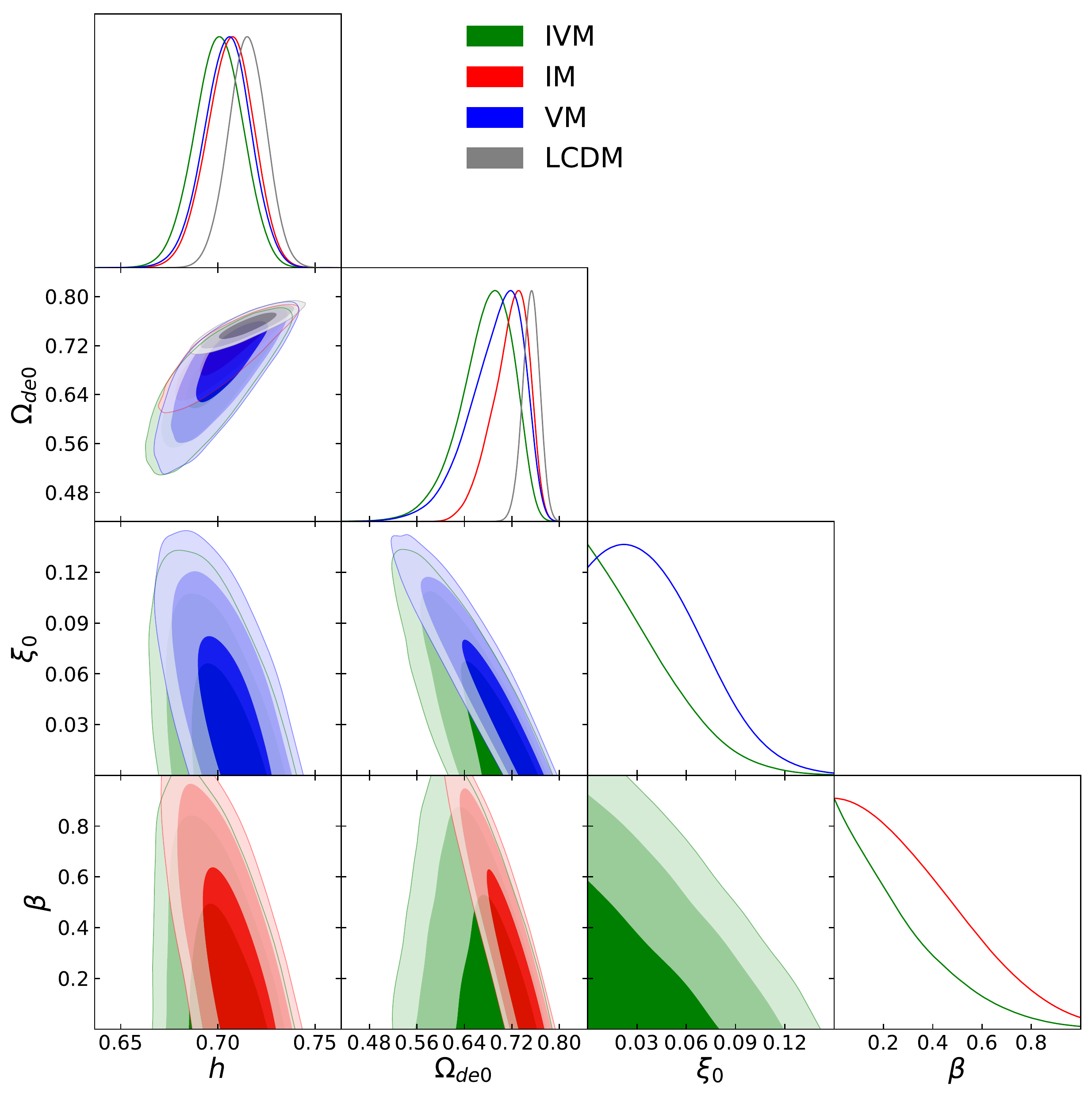} 
\caption{MCMC analysis for the free parameters with 2D contour at $1\sigma$, $2\sigma$, and $3\sigma$ and their 1D posterior distributions for the IVM, IM, VM, and LCDM models.}
\label{fig:contours_hz}
\end{figure}

\begin{figure}
  \centering
  \includegraphics[width=\linewidth]{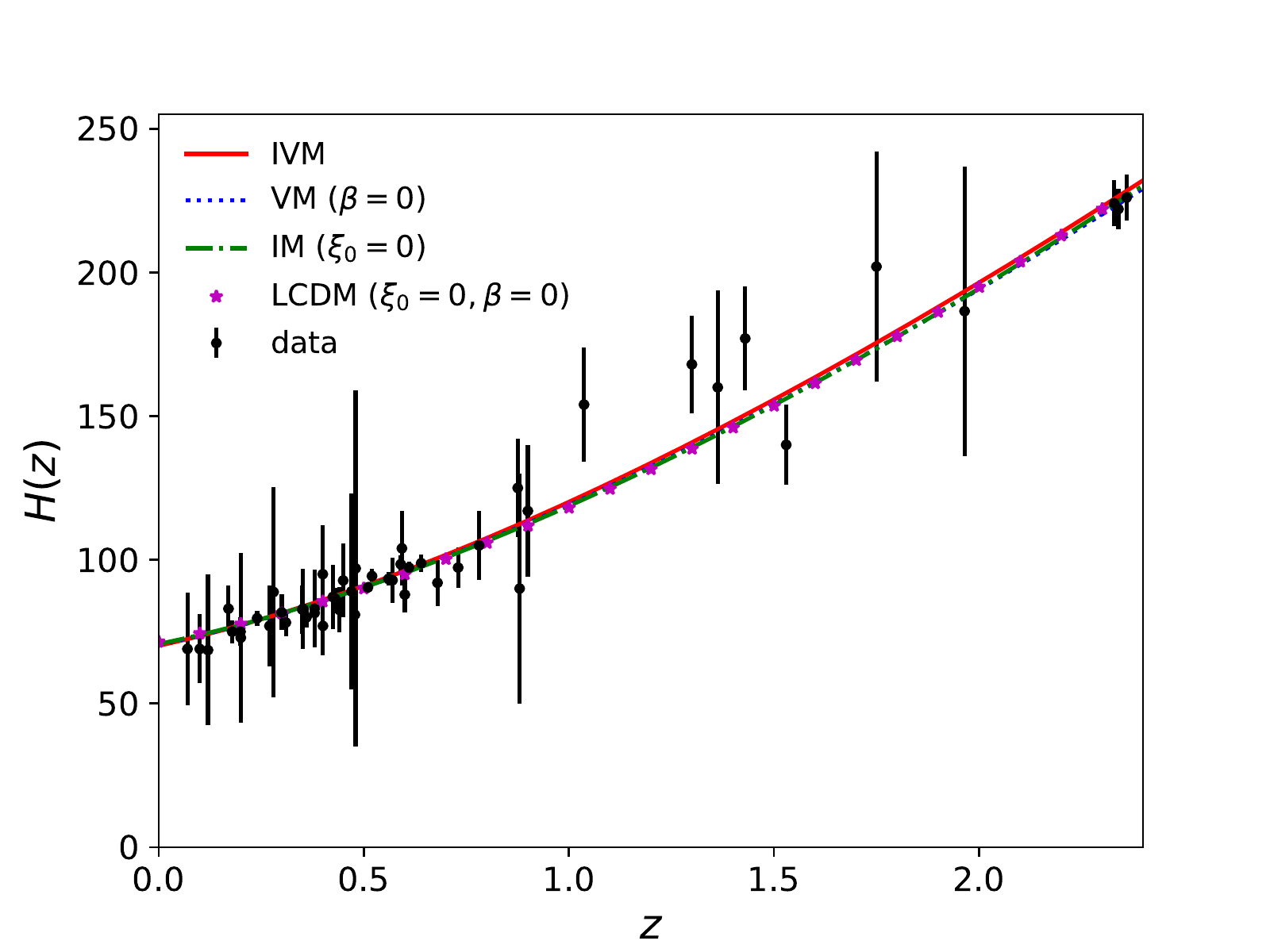} 
\caption{Best fit curves for the IVM (red line), VM (blue dotted line), IM (green dot dashed line), and LCDM (magenta star markers). The black points with uncertainty bars correspond to the OHD sample.}
\label{fig:bf_hz}
\end{figure}

Figure \ref{fig:contours_hz} displays our MCMC analysis for the free parameters with the 2D contours at $68\%$ ($1\sigma$), $95\%$ ($2\sigma$), and $99.7\%$ ($3\sigma$) confidence level (CL) and their corresponding 1D posterior distributions for IVM (green color), IM (red), VM (blue), and LCDM (grey) models. Table \ref{tab:bestfit} shows the best fitting values for the free parameters and their uncertainties at $68\%$ ($1\sigma$) CL of the above mentioned cases. It is interesting to see that $\xi_0$ and $\beta$ are anti-correlated, which is an expected result because both parameters are acting to produce the accelerated expansion of the Universe. On the other hand, with the existence of a viscous Universe or an interacting dark sector (or both), we could establish a upper bound on the $\Omega_{de}$. We will discuss this bound in more detail in the next Section. Finally, Figure \ref{fig:bf_hz} displays the best fit curves over the OHD sample.

\begin{table*}
\caption{Best fit values for the free parameters of IVM, IM, VM and LCDM models using the OHD sample. Additionally, it is reported the $\chi^2$, AIC, BIC, $\Delta$AIC$\equiv$AIC$-$AIC$^{LCDM}$, $\Delta$BIC$\equiv$BIC$-$BIC$^{LCDM}$.}
\centering
\begin{tabular}{| lccccccccc|}
\hline
Model & $\chi^2$ &  $h$ & $\Omega_{de0}$ & $\xi_0$ & $\beta$  & AIC & $\Delta$AIC & BIC & $\Delta$BIC               \\
\hline 
 & & & & & & & & & \\
IVM & $30.5$ & $0.701^{+0.012}_{-0.013}$ &  $0.682^{+0.040}_{-0.040}$ & $0.028^{+0.033}_{-0.020}$ & $0.200^{+0.260}_{-0.145}$ & $38.5$ & $5.6$ & $62.0$ & $17.3$    \\ [1.1ex]
IM & $29.2$ & $0.707^{+0.011}_{-0.012}$ & $0.721^{+0.026}_{-0.037}$ & $0$ & $0.283^{+0.290}_{-0.197}$ & $35.2$ & $2.3$ & $52.8$ & $8.2$   \\ [1.1ex]
VM & $29.1$ & $0.705^{+0.011}_{-0.012}$ & $0.698^{+0.038}_{-0.054}$ & $0.040^{+0.035}_{-0.026}$ & $0$ & $35.1$ & $2.2$ & $52.7$ & $8.1$  \\ [1.1ex]
LCDM & $28.9$ & $0.715^{+0.010}_{-0.010}$ & $0.753^{+0.014}_{-0.015}$ & $0$ & $0$ & $32.9$ & $0$ &  $44.6$ & $0$    \\ [1.1ex]
\hline
\end{tabular}
\label{tab:bestfit}
\end{table*}

%%%%%%%%%%%%%%%%%%%%%%%%%%%%%%%%%%%%%%%%%
\section{Results and discussions} \label{sec:Res}
%%%%%%%%%%%%%%%%%%%%%%%%%%%%%%%%%%%%%%%%%

\begin{figure}[H]
  \centering
  \includegraphics[width=.90\linewidth]{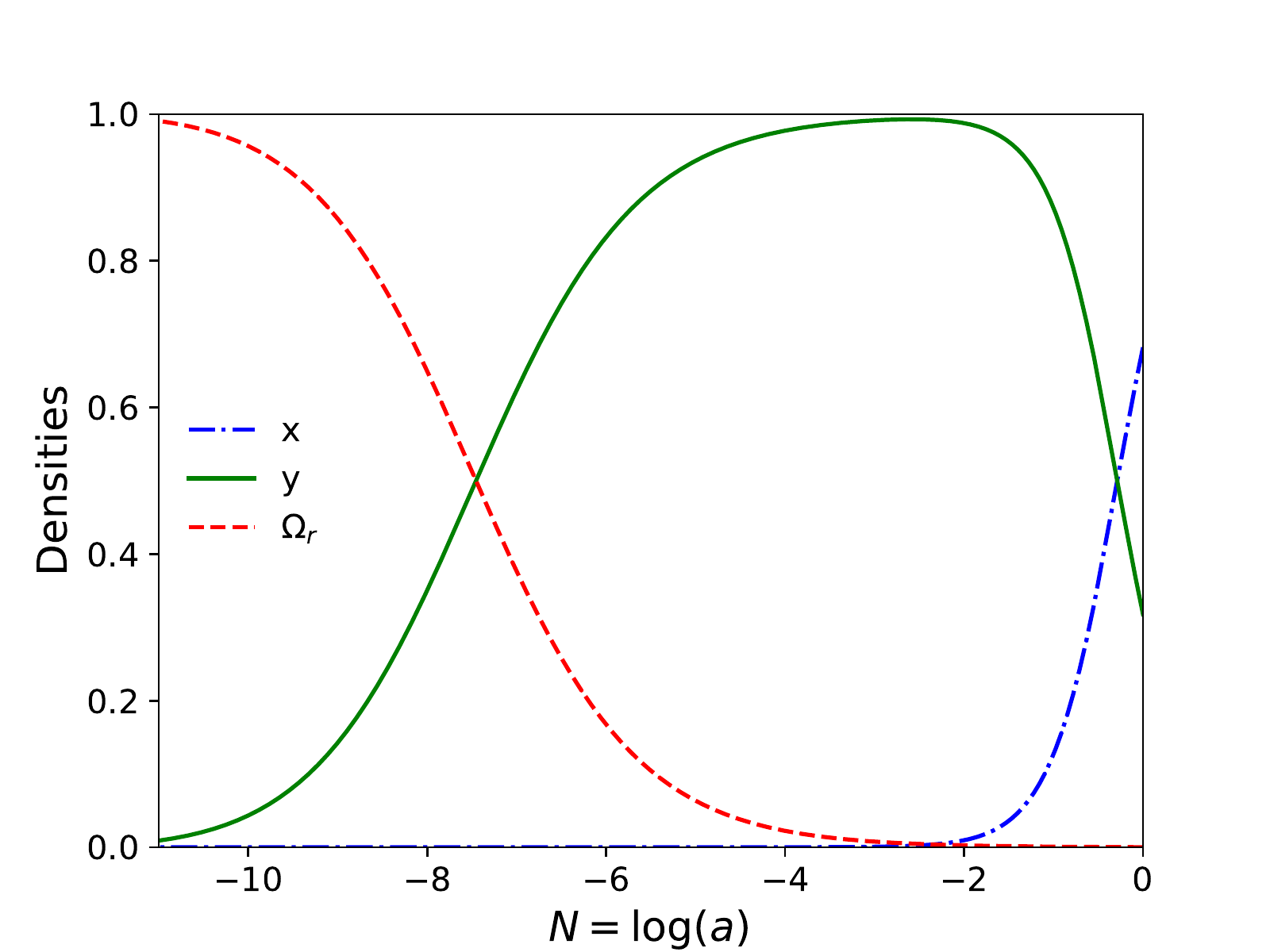} \\
  \includegraphics[width=.90\linewidth]{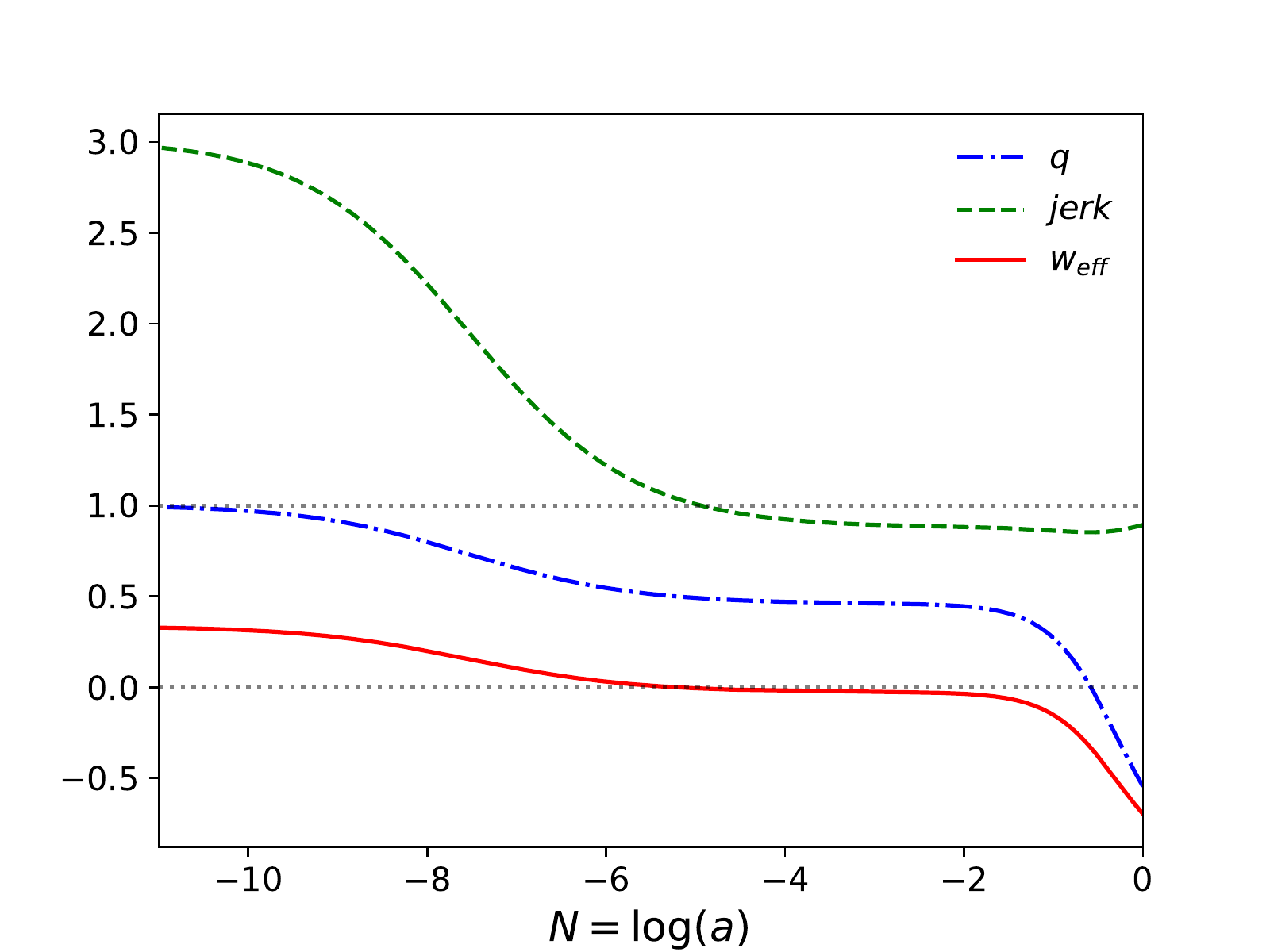} 
\caption{Top panel: Evolution of the dynamical variables $x$, $y$, and $\Omega_r$ for the IVM. Bottom panel: Evolution of the deceleration (blue dot-dashed line) and jerk (green dashed line) parameters and of the effective EoS (red solid line).}
\label{fig:xyN}
\end{figure}

In this section we describe the physical properties of the Universe based on our results of the Bayesian MCMC analysis shown in Table \ref{tab:bestfit}. Figure \ref{fig:xyN} shows the evolution of the dynamical components $x=\Omega_{de}(N)$, $y=\Omega_{dm}(N)$, and $\Omega_r$ of the Universe described by IVM (top panel), and the evolution of the $q(N)$, $j(N)$, and $w_{eff}(N)$ parameters (bottom panel). It is important to remark that the $w_{eff}$ behaves in concordance with standard cosmological model predictions (with Planck data, $w_{eff}^{LCDM}\sim-0.68$ at $z=0$ \cite{Aghanim:2018}). In other words, the Universe is in the quintessence region at late epochs ($-2\lesssim N<0$), 
as dust matter ($w_{eff}\approx 0$) around $-6\lesssim N \lesssim 2$, and takes values closer to $w_{eff}\sim0.3$ in the radiation phase ($N \lesssim -6$). In addition, the deceleration parameter is $q \approx 1/2$ in the dm epoch, and increase towards $q\rightarrow 1$ in the radiation epoch. On the other hand, the jerk parameter $j$ is slightly below the value expected from LCDM ($j=1$) in the region going from the current epoch up to $N\approx -5$, and takes the expected value ($j\rightarrow 3$) for $N<-10$ in the radiation dominated epoch. In summary, the presented models successfully reproduce all the expected epochs and are in good agreement with the LCDM model. 

For a better statistical assessment of different models with different degrees of freedom we use, besides $\chi^2$, the following criteria. The Akaike information criterion (AIC) \cite{AIC:1974, Sugiura:1978} and Bayesian Information criterion (BIC) \cite{schwarz1978} are defined as AIC$\equiv\chi^2 + 2k$ and BIC$\equiv\chi^2+2k\log(N)$ respectively, where $\chi^2$ is the chi-squared function, $k$ is the number of degree of freedom and $N$ is the total number of data, being the model with the lowest value the one preferred by the data (see table III). If the difference in AIC value between a given model and the best one, $\Delta$AIC, is less than $4$, both models are equally supported by the data. For the range $4<\Delta$AIC$<10$, the data still support the given model but less than the preferred one. For $\Delta$AIC$>10$, the observations do not support the given model. Thus, as it is shown in Table \ref{tab:bestfit}, when we compare the IM or VM with respect to LCDM (the preferred model), these models are equally preferred by the data (OHD+SNIa), being the IVM the least preferred by data. Similarly, the difference between a model and the best one, $\Delta$BIC, is interpreted as evidence against a candidate model being the best model. If $\Delta$BIC$<2$, there is no appreciable evidence against the model. In the range $2<\Delta$BIC$<6$, there is a modest evidence against the candidate model and if $6<\Delta$BIC$<10$, the evidence against the candidate model is strong; if $\Delta$BIC$>10$, the evidence against is even stronger. Thus, we have a strong evidence against IM and VM, and even stronger for IVM.

We estimate the deceleration-acceleration transition redshift takes place at $z_T =0.83^{+0.06}_{-0.06}$, $0.82^{+0.05}_{-0.05}$, $0.83^{+0.05}_{-0.05}$, and $0.83^{+0.05}_{-0.05}$ for the IVM, VM, IM, and LCDM respectively, where the uncertainties correspond at $68\%$ CL.  These values are consistent with those reported by \cite{Moresco:2016mzx} of  $z_T^{LCDM} = 0.64^{+0.11}_{-0.06}$ within $1.5\sigma$. On the other hand, they are consistent  with the one obtained by \cite{Kremer:2012} ($z_T \sim 0.74$) when an interacting viscous model is considered. Additionally, our results are also compatible (within $1.8\sigma$ and $1.1\sigma$ respectively) with those found by  Ref. \cite{Jesus:2019} using the non-parametric Gaussian Process method with the OHD data, $z_T = 0.59^{+0.12}_{-0.11}$,and SNIa Pantheon sample, $z_T=0.683^{+0.110}_{-0.082}$, and also the one value found by \cite{Haridasu:2018}, $z_T=0.64^{+0.12}_{-0.09}$ when perform an extension of the standard Gaussian Process to  Supernovae Type-Ia, BAO and Cosmic Chronometers data.

\begin{figure}
  \centering
  \includegraphics[width=\linewidth]{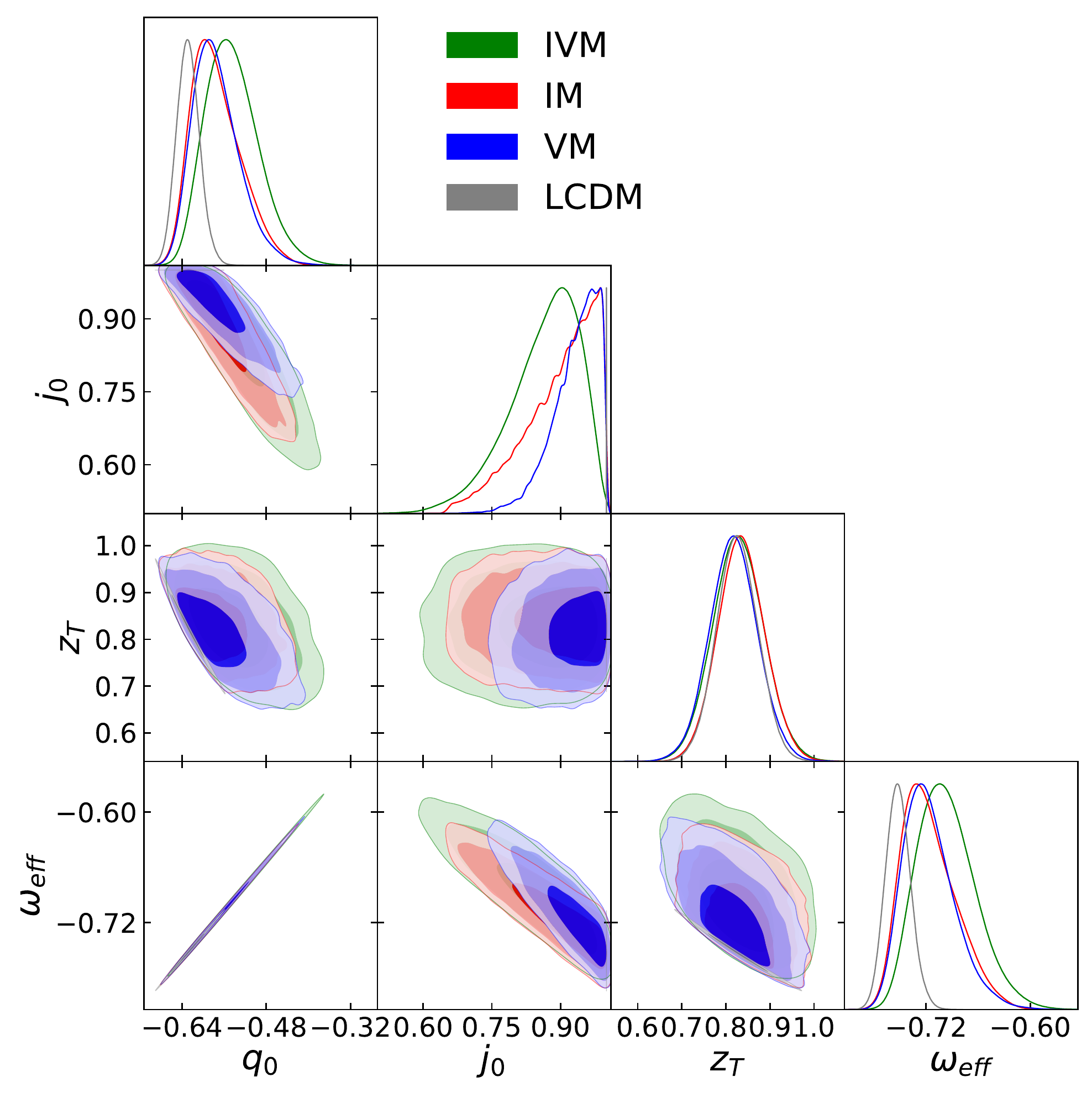} 
\caption{1D posterior distribution of the deceleration and jerk parameters, effective EoS, and $z_T$  and 2D contours at $1\sigma$, $2\sigma$, and $3\sigma$ CL for IVM (green), IM (red) VM (blue), and LCDM (gray).}
\label{fig:contour_q0j0}
\end{figure}

Regarding the cosmographic parameters, we obtain the deceleration one at $z=0$ to be 
$q_0 = -0.55^{+0.06}_{-0.05}$, 
$-0.58^{+0.05}_{-0.04}$, 
$-0.58^{+0.05}_{-0.05}$, 
and $-0.63^{+0.02}_{-0.02}$ 
for the IVM, VM, IM, and LCDM respectively. When we compare these results with the one obtained by \cite{Garcia-Aspeitia:2018fvw} for the LCDM model, $q_0^{LCDM} = -0.54 \pm 0.07$, we find a deviation within $1.3\sigma$. Additionally, their corresponding value of the jerk parameter are 
$j_0 = 0.87^{+0.06}_{-0.09}$, 
$0.94^{+0.04}_{-0.06}$, 
$0.91^{+0.06}_{-0.10}$, 
and $1.0$. On the other hand, when we compare our $q_0^{VM}$ and $j_0^{VM}$ values with those obtained by \cite{Almada:2019} considering viscous models, we find a deviation of about $1.3\sigma$. Additionally, we find a deviation on $q_0$ within $1.2\sigma$ for IVM, IV, and VM to the one value obtained in \cite{Haridasu:2018}. Figure \ref{fig:contour_q0j0} shows 1D posterior distribution and 2D contours of the $q_0$, $j_0$, $w_{eff}(z=0)$, and $z_T$ for IVM, IM, VM, and LCDM. Based on the IVM, it is noteworthy that $w_{eff}$ presents a positive correlation ($corr>0.99$) with $q_0$
and a negative one ($corr=-0.87$) with $j_0$. 
The deceleration-acceleration transition has a negative correlation ($corr=-0.45$) with $q_0$ and a negligible correlation with $j_0$. Between the cosmographic parameters ($q_0,j_0$), we find a negative correlation of $corr=-0.45$. 

\begin{figure}
  \centering
  \includegraphics[width=\linewidth]{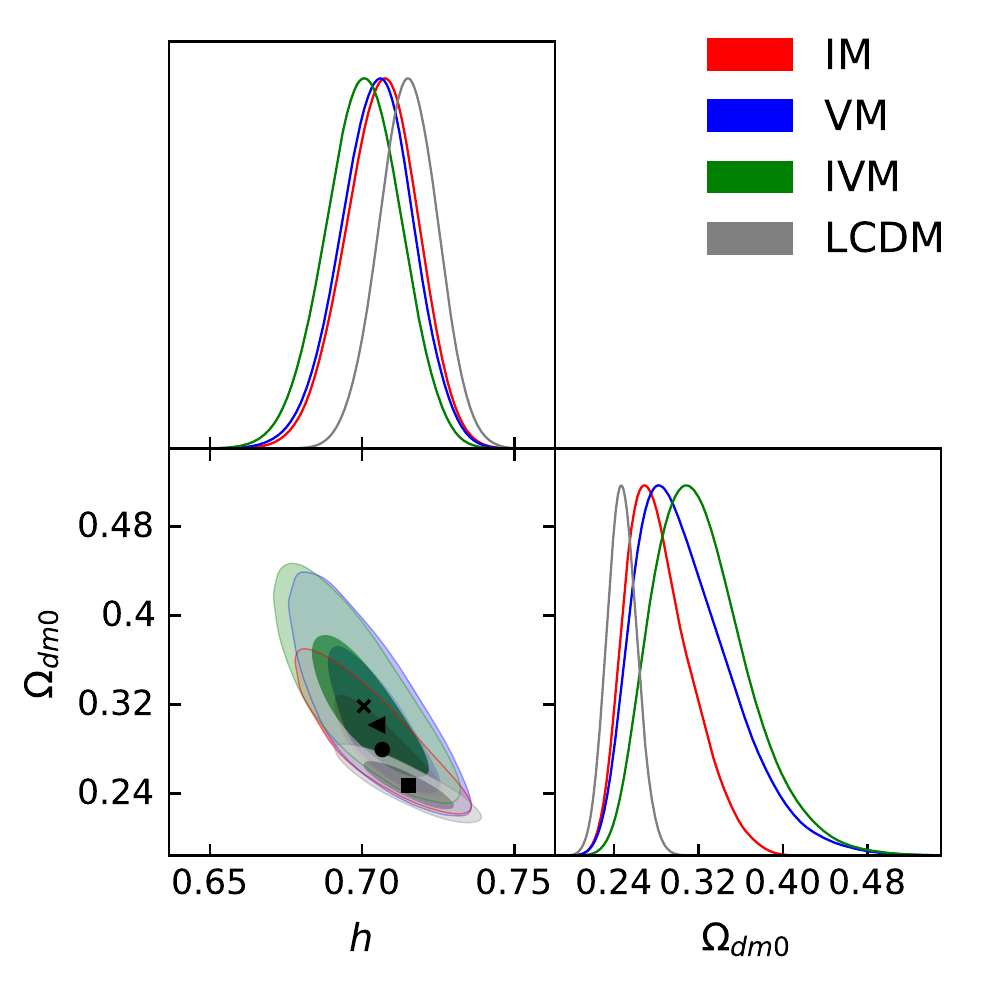} 
\caption{1D posterior distribution of the model parameters and 2D contour on the plane $h$ vs $\Omega_{dm0}$ at $1\sigma$ (darker color) and $2\sigma$ (lighter color) CL for IVM (green), IM (red) VM (blue), and LCDM (gray). The best fit values of $h$ and $\Omega_{dm0}$ are represented by cross (IVM), triangle (VM), circle (IM), and square (LCDM) markers.}
\label{fig:contour_hOdm}
\end{figure}

Figure \ref{fig:contour_hOdm} displays the 1D posterior distribution of the variables $h$ and $y_0=\Omega_{dm0}$ for the IVM (green), IM (red), VM (blue), and LCDM (gray) together with the $h$-$\Omega_{dm0}$ contour at $1\sigma$ and $2\sigma$ CL. It is interesting to see that a possible effect of the interaction and viscosity terms is to increase the dm component at current epochs; nevertheless, such contributions are consistent within $2\sigma$ CL to the value of LCDM.

Only considering the DM component as the viscous one, the authors in \cite{Velten:2013} argue that, in the presence of several fluids in the Universe, it is difficult to distinguish which fluid is producing the viscous effects at the background level, in other words, there is degeneracy. However, as we mentioned in the introduction, when the DM fluid is the responsible for such dissipative effects it means that, at $z<1$, the DM particles probably do not decay into energetic relativistic particles, such as sterile neutrinos or supersymmetric DM \cite{Wilson:2007}. In this context, the DM particles should have a mass of the order $\sim 1 {\rm MeV}$ and a lifetime of order the Hubble time $H_0^{-1}$. Based on the expression $\zeta = 1.25 \rho_h \tau_e [ 1 - (\rho_l + \rho_{r})/\rho ]^2$ presented in \cite{Wilson:2007}, where $\rho$ is the total energy density in the Universe, $\rho_h$ is the DM density for an unstable decaying DM, $\rho_l$ is the produced relativistic energy density, and $\tau_e = \tau/(1-3H\tau)$ is the equilibrium time with $\tau$ being the particle decaying time. Hence, we could give a bounded relation for such densities and the decaying lifetime at $z=0$ as  $1.25 \kappa^2 \rho_{h0} \tau_e [ 1 - (\rho_{l0} + \rho_{r})/\rho_{0} ]^2\, / H_0 \sqrt{\Omega_{dm0}} <0.086, 0.098$ at $95\%$ CL for the IVM and VM, respectively.

%%%%%%%%%%%%%%%%%%%%%%%%%%%%%%%%%%%%%%%%
\section{Conclusions and Outlooks} \label{sec:Con}
%%%%%%%%%%%%%%%%%%%%%%%%%%%%%%%%%%%%%%%%%

In this work we have addressed a phenomenological model for a flat Universe containing a radiation component and a viscous fluid (dark matter plus baryons) that interacts with a perfect fluid (DE), denoted as IVM. The IVM is characterized by the parameter phase-space ${\bf \Theta}= (h, \Omega_{de0}, \xi_0, \beta)$. Furthermore, we studied some particular cases of the model by considering the dm fluid as an interacting perfect fluid ($\xi_0=0$), and as only viscous fluid ($\beta=0$). The latter consisted of the non-interacting perfect fluid ($\beta=\xi_0=0$) which corresponds to the LCDM model. In the first part of the work, we studied the IVM from a dynamical approach. We obtained the stability conditions for the critical points presented in the Table \ref{tab:CP}, which are in concordance with those obtained by \cite{Leyva:2017} when a linear interacting term of the form $z(x)=\alpha x$ is considered, being $\alpha$ an appropriate constant. Figure \ref{fig:phasespace} shows the phase-space of the dynamical system where the color gradient represents the value of the deceleration parameter ($q$), jerk ($j$), and effective EoS ($w_{eff}$).
The second part of the work consisted in performing a Bayesian MCMC analysis using the largest sample of the Hubble parameter data to obtain the best fit parameters for each model (Table \ref{tab:bestfit}). Then, we reconstructed the deceleration and jerk parameter and the effective EoS, as was shown in Fig. \ref{fig:xyN}. We estimate the current values of the cosmographic parameters as 
$q_0 = -0.55^{+0.06}_{-0.05}$, 
$-0.58^{+0.05}_{-0.04}$, 
$-0.58^{+0.05}_{-0.05}$, 
$-0.63^{+0.02}_{-0.02}$
and
$j_0 = 0.87^{+0.06}_{-0.09}$, 
$0.94^{+0.04}_{-0.06}$, 
$0.91^{+0.06}_{-0.10}$,
$1.0$,
for the IVM, VM, IM, and LCDM respectively, which are in agreement with those reported in the literature  considering other models \cite{Garcia-Aspeitia:2018fvw, Almada:2019}. Finally, although our results on BIC suggest the models used are unfavourable over LCDM standard paradigm, they give an alternative to alleviate the CC problems by adding some degree of freedom to LCDM. 

\section*{Acknowledgments}
We thank the anonymous referee for thoughtful remarks and suggestions. The authors acknowledge the enlightening conversation with Jes\'us Astorga. M.A.G.-A. acknowledges support from SNI-M\'exico, CONACyT research fellow, COZCyT and Instituto Avanzado de Cosmolog\'ia (IAC) collaborations. J.M. acknowledges the support from CONICYT project Basal AFB-170002, V.M. acknowledges the support of Centro de Astrof\'{\i}sica de Valpara\'{\i}so (CAV). J.M., M.A.G.-A. and V.M. acknowledge CONICYT REDES (190147). \\

\bibliographystyle{unsrt}
\bibliography{librero0}

% Don't change these lines
%\bsp	% typesetting comment
%\label{lastpage}

\end{document}